# Emergent zero-field anomalous Hall effect in a reconstructed rutile antiferromagnetic metal

Meng Wang[1,*,†], Katsuhiro Tanaka[2,†], Shiro Sakai[1], Ziqian Wang[1], Ke Deng[3,4], Yingjie Lyu[5], Cong Li[5], Di Tian[5], Shengchun Shen[6], Naoki Ogawa[1,7], Naoya Kanazawa[8], Pu Yu[5], Ryotaro Arita[1,2], Fumitaka Kagawa[1,9,*]

[1]*RIKEN Center for Emergent Matter Science (CEMS), Wako, 351-0198, Japan.*
[2]*Research Center for Advanced Science and Technology, University of Tokyo, Tokyo 153-8904, Japan.*
[3]*Shenzhen Institute for Quantum Science and Engineering, Southern University of Science and Technology (SUSTech), Shenzhen 518055, China.*
[4]*International Quantum Academy, Shenzhen 518048, China.*
[5]*State Key Laboratory of Low Dimensional Quantum Physics and Department of Physics, Tsinghua University, Beijing, 100084, China.*
[6]*Department of Physics, University of Science and Technology of China, Hefei, 230026, China.*
[7]*Department of Applied Physics and Quantum-Phase Electronics Center (QPEC), University of Tokyo, Tokyo 113-8656, Japan.*
[8]*Institute of Industrial Science, The University of Tokyo, Tokyo 153-8505, Japan.*
[9]*Department of Physics, Tokyo Institute of Technology, Tokyo 152-8551, Japan.*
[*]Corresponding authors. Email: meng.wang@riken.jp; fumitaka.kagawa@riken.jp
[†]These authors are equally contributed.

**Anomalous Hall effect (AHE) emerged in antiferromagnetic metals shows intriguing physics and application potential. In contrast to certain noncollinear antiferromagnets, rutile $RuO_2$ has been proposed recently to exhibit a crystal-assisted AHE with collinear antiferromagnetism. However, in $RuO_2$, the on-site magnetic moment accompanying itinerant *4d* electrons is quite small, and more importantly, the AHE at zero external field is prohibited by symmetry because of the high-symmetry [001] direction of the Néel vector. Here, we show the AHE at zero field in the collinear antiferromagnet, Cr-doped $RuO_2$. The appropriate doping of Cr at Ru sites results in a rotation of the Néel vector from [001] to [110] and enhancement of the on-site magnetic moment by one order of magnitude while maintaining a metallic state with the collinear antiferromagnetism. The AHE with vanishing net moment in the $Ru_{0.8}Cr_{0.2}O_2$ exhibits an orientation dependence consistent with the [110]-oriented Néel vector. These results open a new avenue to manipulate AHE in antiferromagnetic metals.**

## Introduction

Anomalous Hall effect (AHE) had been considered for a long time as a unique feature of ferromagnetic metals, and its magnitude was empirically taken as proportional to the macroscopic magnetization $M$ [1,2]. It follows that in antiferromagnetic materials, which host zero macroscopic magnetization or only small canting moments, the AHE should be negligibly small. However, modern theories indicate that in some antiferromagnetic materials, the AHE can be expected if the magnetic space group (MSG) (or, equivalently, the magnetic point group that the MSG belongs to) allows for a nonzero Berry curvature and/or asymmetric scattering, even if the corresponding macroscopic magnetization is zero.[3-5] Such an AHE has been experimentally demonstrated for various noncollinear antiferromagnets with magnetic multipoles [3-12], such as kagome $Mn_3Sn$ and pyrochlore $R_2Ir_2O_7$.

From the symmetry point of view, an antiferromagnetism-induced AHE can also be expected in a collinear antiferromagnet. A prototypical candidate material that has been extensively considered is the rutile antiferromagnetic $RuO_2$ [14-18]. As shown in **Fig. 1a**, the crystal structure of $RuO_2$ consists of two Ru sublattices with antiparallel magnetic moments. The two magnetic sublattices have different chemical environments due to the asymmetric O-Ru-O bond configuration. The simplest argument to determine the presence or absence of the AHE under collinear antiferromagnetism would be to consider how the Hall vector $\boldsymbol{\sigma}_{Hall} = (\sigma_{yz}, \sigma_{zx}, \sigma_{xy})$ is transformed by the symmetry operations. For instance, if the Néel vector ($\boldsymbol{L}$) of $RuO_2$ is along the [110] direction, the MSG is $Cmm'm'$, in which $\boldsymbol{\sigma}_{Hall}$ along [110] is invariant under all symmetry operations and thus allows for a zero-field AHE [14]. In contrast, if $\boldsymbol{L} \parallel [001]$, the MSG is $P4'_2/mnm'$ [14], which does not allow for a finite $\boldsymbol{\sigma}_{Hall}$ because no vector can be invariant under two orthogonal rotation symmetry operations (see Supplementary Note 1 for details). A previous neutron experiment indicates that the Néel vector in $RuO_2$ is along [001],[15] and hence $\boldsymbol{\sigma}_{Hall}$ and the zero-field AHE are prohibited by symmetry (Supplementary Fig. 1). To unveil the AHE associated with the collinear antiferromagnetism in $RuO_2$, a recent study focused on tilting the Néel vector from [001] toward [110] by utilizing a high magnetic field of ~50 T. [17,18] This phenomenon can be viewed as a magnetic-field-induced AHE associated with a Néel vector, forming a sharp contrast to AHEs in ferromagnets, in which the AHE can be observed even under zero-field. Thus, achieving a zero-field AHE in such a rutile-type collinear antiferromagnet remains challenging in experiments.

The previous density functional theory (DFT) calculations have revealed that the easy axis of the Néel vector in $RuO_2$ sensitively depends on the electron filling, [17] which inspired us to pursue the zero-field AHE in the derivatives of $RuO_2$ by means of appropriate modulations on its Fermi level. To change the direction of the Néel vector from [001] and render the zero-field AHE allowed by symmetry, we dope Cr into $RuO_2$. Note that the $4d$ orbital level of $Ru^{4+}$ is slightly higher than the $3d$ orbital level of $Cr^{4+}$, a charge transfer from $Ru^{4+}$ to $Cr^{4+}$ ions can naturally be expected (**Fig. 1b**)[19,20] while favouring anti-parallel spin coupling between the nearest-neighbouring Ru and Cr sites. Besides, considering that collinear spin orders are realized in both $RuO_2$ (antiferromagnetic) and $CrO_2$ (ferromagnetic) in rutile phases [15,21,22], the collinear antiferromagnetic state is reasonably expected in stoichiometric proximity to $RuO_2$. In this work, our magnetometry confirms that the direction of the Néel vector in the $Ru_{0.8}Cr_{0.2}O_2$ film is driven to the [110] direction. Concomitantly, we find that the $Ru_{0.8}Cr_{0.2}O_2$ film exhibits an appreciable zero-field AHE with hysteretic behaviour while the net magnetization is vanishingly small.

## Results

**DFT+DMFT calculations on the impact of Cr-doping.**

To gain insight into the impact of Cr-doping on the Fermi level, we first performed DFT calculations for the paramagnetic states of $Ru_{1-x}Cr_xO_2$ for x = 0 and 0.5. As shown in **Fig. 1c**, by doping Cr, the shift of the projected density of states (or, equivalently, the shift of the Fermi level) is observed, as expected. The magnetic calculation for x = 0.5 (**Fig. 1d**) further demonstrates that the ground state has appreciable local magnetic moments with antiparallel couplings among the nearest neighboring Cr and Ru ions. Note that the DFT+$U$ calculations on $RuO_2$ show that the energy difference with the Néel vector orienting to [001], [100], and [110] is tiny (~ 5 meV) and that the easy-axis direction sensitively depends on the Fermi level (Supplementary Fig. 2). [17] Our DFT results therefore support our working hypothesis that Cr doping is a promising approach to change the Néel vector direction while maintaining the collinear antiferromagnetic order.

The DFT results indicate that Cr doping is also accompanied by the enhancement of the local magnetic moment. For the case of non-doped $RuO_2$, the Ru ions exhibit a negligibly small spin polarization when $U$ is small (< 1 eV) (Supplementary Fig. 3). Such a small on-site moment is ascribed to the itinerant $4d$ orbital, presumably consistent with the quite small moment (~0.05 $\mu_B$ per site) observed by neutron experiment in $RuO_2$. In contrast, when Cr is doped, considerable local moments are observed (0.15 $\mu_B$ for x = 0.25 and 0.4 $\mu_B$ for x = 0.5; see Supplementary Fig. 3) in the DFT calculations, even at $U = 0$.

Thus, based on our DFT calculations, we can expect that (i) the easy axis of the Néel vector changes from the original [001] direction, in which the zero-field AHE is prohibited, to another direction, and (ii) the impact of the collinear antiferromagnetic ordering on the transport properties is more observable due to the enhancement of the local magnetic moments. These expectations should be verified by the experiments below.

**Films fabrication and valence evaluation.**

We synthesized the $Ru_{1-x}Cr_xO_2$ films by pulsed laser deposition (PLD) on $TiO_2$ (110) substrates with x =0.1, 0.2, and 0.3 (see Methods). The high crystalline quality of the films was confirmed by X-ray $2\theta$-$\omega$ scans (see supplementary Fig. 4a) and the surface topography with atomic terraces (Supplementary Fig. 4c). Besides, the resistivities of the materials increase as the doping level increases, while all compounds show a metallic behavior, as shown in Supplementary Fig. 5a. The robust metallicity implies the strong overlap of Cr and Ru orbitals.

To probe the valence state of the doped Cr in the rutile lattice, we carried out soft X-ray absorption spectroscopy (XAS) measurements (see Methods) on the three films. **Figure 2a** shows the XAS results near the *L*-edge of Cr, with a comparison to that from $La_{1-x}Sr_xCrO_3$ materials[23]. The Cr in all of the $Ru_{1-x}Cr_xO_2$ films exhibits a fractional valence state between +3.25 and +3.5. As the doping level increases from 0.1 to 0.3, the peak shows a gradual shift to lower energy, indicating a gradual decrease in valence. Such a tendency is consistent with our scenario that the Cr doping is accompanied by the charge transfer and the corresponding Fermi-level shift.

**Antiferromagnetic metal phases in the $Ru_{1-x}Cr_xO_2$ films.**

To check whether the magnetic ground state is still antiferromagnetic upon the Cr doping, we performed magnetic susceptibility ($\chi$) and magnetization ($M$) measurements with magnetic field ($H$) and temperature ($T$) dependences (see Methods and Supplementary Fig. 6 for details). The results are summarized in **Figs. 2b, c**, and we first focus on the results of x = 0.1 and 0.2. The high-temperature regions of the $\chi^{-1}$–$T$ profiles are fitted with the Curie–Weiss law, $\chi = C/(T-\theta_W)$, and we obtain $\theta_W \approx$ −10 K and −75 K for x = 0.1 and 0.2, respectively. These results indicate that an antiferromagnetic interaction is dominant in x = 0.1 and 0.2 [24-27]. Moreover, compared with the local moment of ~0.05 $\mu_B$ per site in pure $RuO_2$,[15] the effective on-site moments ($\mu_{eff}$) obtained from the fittings are distinctly enhanced in x = 0.1 (~0.9 $\mu_B$ per site) and x = 0.2 (~2.1 $\mu_B$ per site) (Fig. 2b, inset and Supplementary Note 2).[24,25] This pronounced enhancement is also consistent with our DFT calculations.

The *M*–*H* curves at the lowest temperature, 3 K, demonstrate that the spontaneous net magnetization at zero field is too small to be distinguished in the antiferromagnetic $Ru_{0.9}Cr_{0.1}O_2$ and $Ru_{0.8}Cr_{0.2}O_2$ (**Fig. 2c**). Moreover, the field-induced moment at 7 T is only 0.03 $\mu_B$ (x = 0.1) and 0.04 $\mu_B$ (x = 0.2) per formula unit ($\mu_B$/f.u.), which are almost two orders of magnitude smaller than that in ferromagnetic $SrRuO_3$ [28,29] and $CrO_2$ [30], excluding the possibility of a ferromagnetic ground state for x = 0.1 and 0.2. In addition, a Kerr mapping was also carried out in the $Ru_{0.8}Cr_{0.2}O_2$ film by utilizing a high-resolution equipment at 7 K and 0 T (see Supplementary Fig. 7).

However, the observed Kerr rotation (~μrad) is three orders of magnitude smaller than that (~mrad) in the ferromagnetic SrRuO$_3$ film [31], and no domain walls can be observed, which further indicates that an antiferromagnetic state is preserved with a vanishingly small net magnetization at zero magnetic field.

In the Ru$_{0.7}$Cr$_{0.3}$O$_2$ film, contrastingly, the analysis based on the Curie–Weiss law results in a small positive $\theta_W$ with $\mu_{eff}$ of ~2.5 $\mu_B$ per site (Fig. 2b, and Supplementary Note 2). Furthermore, the *M–H* curve exhibits a finite remanent magnetization, and the magnetization at 7 T is distinctly larger compared with the case of x = 0.1 and 0.2. These observations indicate the evolution of a ferrimagnetic phase in x = 0.3, consistent with the tendency from RuO$_2$ to CrO$_2$ [15,21]. Therefore, the AHE accompanying the ferrimagnetic phase in x = 0.3 is beyond the scope of this study.

**Néel-vector direction in the Ru$_{0.8}$Cr$_{0.2}$O$_2$ film.**

We then focus on the antiferromagnetic Ru$_{0.8}$Cr$_{0.2}$O$_2$ (110) sample, which exhibits a pronounced $\mu_{eff}$ of ~2.1 $\mu_B$ per site, and aim to reveal the direction of the Néel vector. The DFT calculations in RuO$_2$ suggest a finite net magnetic moment when the Néel vector along [100] is assumed (Supplementary Fig. 2a), which should be preserved in the doped phase. Our *M-H* measurements in Ru$_{0.8}$Cr$_{0.2}$O$_2$ show a vanishing net moment, thereby ruling out the possibility that the Néel vector is along [100]. Then, the remaining candidates of the Néel-vector direction are the [001] and [110] orientations. To test these two possibilities, we refer to the fact that the field-induced moment in a collinear antiferromagnet is generally minimized when the field is parallel to the Néel vector, as illustrated in **Fig. 3** inset [17,25]. The anisotropy of the field-induced moment was measured on the Ru$_{0.8}$Cr$_{0.2}$O$_2$ (110) film for the fields of the out-of-plane [110] and in-plane [001] directions. The anisotropic response demonstrates that the [110] axis exhibits a smaller field-induced moment (Fig. 3) and thus the Néel vector should be along [110], rather than [001], in Ru$_{0.8}$Cr$_{0.2}$O$_2$. The corresponding MSG is *Cmm'm'*, and hence the zero-field AHE is allowed by symmetry.[14,16]

**AHE in the $Ru_{0.8}Cr_{0.2}O_2$ (110) film.**

The longitudinal resistivity and Hall conductivity in $Ru_{0.8}Cr_{0.2}O_2$ (110) film were measured with currents along two in-plane directions, [001] and $[1\bar{1}0]$, as shown in Supplementary Fig. 5 (see Methods). Both directions show a metallic state, and the Hall conductivity measured with the current along $[1\bar{1}0]$ exhibits a larger signal. Therefore, we below present the results of the AHE with the current along $[1\bar{1}0]$.

**Figure 4a** shows the Hall conductivity ($\sigma_{xy}$) with a magnetic field sweeping at 3 K. Distinctly, a hysteretic feature is observed, in stark contrast to the absence of a hysteretic behavior in the *M-H* curve (Fig. 2c). This behavior demonstrates that the finite Hall vector is involved in the $Ru_{0.8}Cr_{0.2}O_2$ (110) film, even though the net magnetization is vanishingly small within the experimental accuracy. Thus, in the magnetic field range in which $\sigma_{xy}$ shows hysteretic behavior, one should take into account the coexistence of the two magnetic domains with opposite Hall vectors (i.e., the AHCs with opposite signs).

In general, the origin of $\sigma_{xy}$ consists of the external magnetic field (or ordinary Hall conductivity, $\sigma_{xy}^{OHE}$, proportional to $H$ with a coefficient $k_o$) and the magnetism (or anomalous Hall conductivity, $\sigma_{xy}^{AHE}$). The $\sigma_{xy}^{AHE}$ is often dictated by the contribution proportional to the net magnetization, but in the present system, the antiferromagnetic order coupled with the special lattice symmetry can also contribute [3,6,14]. Thus, the observed $\sigma_{xy}$ can be described as the sum of the three components:

$$\sigma_{xy}(H) = \sigma_{xy}^{OHE}(H) + \sigma_{xy}^{M}(H) + \sigma_{xy}^{AF}(H)$$
$$= k_o \cdot H + k_m \cdot M(H) + \sigma_{xy}^{AF}(H), \qquad (1)$$

where $\sigma_{xy}^{M}$ is the anomalous Hall conductivity proportional to the field-induced net magnetic moment $M$ with a coefficient $k_m$, and $\sigma_{xy}^{AF}$ is the anomalous Hall conductivity arising from the antiferromagnetic ordering [6]. Note that in the present field range, the magnetic field-dependent $\sigma_{xy}^{AF}(H)$ is caused by the change in the relative volume of the two types of antiferromagnetic domains with opposite signs of AHC.

At sufficiently high magnetic fields, the hysteretic behavior disappears, and therefore,

a single antiferromagnetic domain is expected. Thus, $\sigma_{xy}^{AF}$ is considered to be a constant, $\sigma_{xy}^{AF,0}$, at a sufficiently high magnetic field[14]. Utilizing the data of Hall conductivity and magnetization at 4–7 T, where the hysteretic behavior is absent, we can thus obtain the coefficients, $k_o$ and $k_m$, and $\sigma_{xy}^{AF,0}$. For clarity, by subtracting $\sigma_{xy}^{OHE} = k_o \cdot H$, we display the experimental $\sigma_{xy}^{AHE}$ together with the fitting curve $k_m \cdot M + \sigma_{xy}^{AF,0}$ as a function of the net magnetization in **Fig. 4b**. The value of $\sigma_{xy}^{AF,0}$ is ≈3.2 S/cm, which is indicated by the intercept of the fitting curve at $M = 0$. In the low-field region, the experimental $\sigma_{xy}^{AHE}(H)$ deviates from the linear fitting. In the present framework, this deviation is attributable to the coexistence of two antiferromagnetic domains with opposite signs of AHC.

The evolutions of $\sigma_{xy}^{AF}$ and $\sigma_{xy}^{M}$ with magnetic field sweeping at 3 K are shown in **Fig. 4c**, where $\sigma_{xy}^{M}$ is set to $k_m \cdot M$, and $\sigma_{xy}^{AF}$ is obtained by subtracting $\sigma_{xy}^{M}$ from $\sigma_{xy}^{AHE}$. Interestingly, the $\sigma_{xy}^{AF}$ shows a hysteretic profile and a clear remnant value even at the vanishing net moment (Fig. 2c). Such features indicate an AHC contributed by the antiferromagnetic ordering, not due to the canting moment. The emergent $\sigma_{xy}^{AF}$ decreases as the temperature increases and disappears at 40–50 K (**Fig. 4d** and Supplementary Fig. 8), indicating the antiferromagnetic order transition point ($T_N$). Note that the noncollinear antiferromagnetic materials with complicated spin interactions generally show a large value of $|\theta_W/T_N|$ (> 10) [32]. Here, the small value of $|\theta_W/T_N| = 1.5$–1.8 in the Ru$_{0.8}$Cr$_{0.2}$O$_2$ film is typically located in the regime of collinear antiferromagnets.

To gain further insight into the microscopic mechanisms of the $\sigma_{xy}^{AF}$ and $\sigma_{xy}^{M}$, we compared the AHC–$\sigma_{xx}$ scaling curves [2,33-36] among Ru$_{0.8}$Cr$_{0.2}$O$_2$ (110) films with different $\sigma_{xx}$, which was tuned by tailoring the thickness. As shown in Supplementary Fig. 9, all films are located at the crossover from dirty to intermediate regimes with $10^3 < \sigma_{xx} < 10^4$ S/cm, thereby ruling out the skew scattering contribution, which is generally considered in high conductive metals ($\sigma_{xx} > 10^6$ S/cm). Besides, a further analysis based on the $\sigma_{xy}^{M}(T)$–$\sigma_{xx}(T)^2$ profile gives an intrinsic Berry curvature term of 14 S/cm (Supplementary Note 3) and the extrinsic side-jump contribution of ~ 10 S/cm. These results indicate that the Berry curvature and extrinsic scattering microscopic mechanisms both contributes to $\sigma_{xy}^M$ (T) in our films. We note that the $\sigma_{xy}^M$ value is similar to the AHC in ferromagnetic SrRuO$_3$ films grown by PLD, although the canting moment (0.04 μB/f.u.) of our Ru$_{0.8}$Cr$_{0.2}$O$_2$ (110) film is ~40 times smaller than the ferromagnetic moment in SrRuO$_3$ films [37,38]. We also note that the value of $\sigma_{xy}^{AF}$ in Ru$_{0.8}$Cr$_{0.2}$O$_2$ is one order of magnitude larger than the recently reported collinear antiferromagnetic semiconductor MnTe [39].

**Orientation-anisotropic anomalous Hall response.**

Finally, we show that the transport properties in our Ru$_{0.8}$Cr$_{0.2}$O$_2$ film also indicate the Hall vector along [110]. To address this issue experimentally, we referred to the fact that the transverse anomalous Hall current ($J_H$) is given by $\boldsymbol{J}_H = \boldsymbol{E} \times \boldsymbol{\sigma}_{Hall}$, [14,16] where $\boldsymbol{E}$ represents the applied external electric field, and carried out transport measurements on another film grown on TiO$_2$ (100). Herein, the current was applied along the [010] direction to keep the Hall voltage also along the [001] direction for comparison. The temperature dependence of AHE is similar to that observed for the [110]-oriented films (Supplementary Fig. 8), indicating that the transition temperature is not affected by the orientation of the substrate. As shown in **Fig. 5a** and **5b**, the longitudinal conductivities at low temperatures of the two films are very close to each other, while the $\sigma_{xy}^{AHE}$ that emerges from the (100) film is distinctly smaller than the value for the film grown on TiO$_2$ (110). Upon further analyzing the magnetization and the anomalous Hall contributions of $\sigma_{xy}^M$ and $\sigma_{xy}^{AF}$, as shown in **Fig. 5c**, we find that both of the anomalous Hall components are suppressed compared with those in **Fig. 4c**, although the emergent magnetization is increased. Furthermore, we find that the saturated $\sigma_{xy}^{AF}$ in the [100]-oriented film is approximately ~2.2 S/cm, which is 0.7 (≃ sin45°) times that in the [110]-oriented sample, ~3.2 S/cm. Independent of the symmetry arguments based on the Néel-vector direction along [110], these transport results further support that the Hall vector is directed along the [110] direction in this compound, as illustrated in Fig. 5c, inset.

In summary, by tuning the 3*d*-4*d* orbital reconstruction to achieve symmetry manipulation and balance the itinerant properties and the electron correlation, we have succeeded in observing the zero-field AHE in the collinear antiferromagnetic rutile metal. Note that the antiferromagnetic metallic phase is extremely rare in correlated oxides [26,27,40], and such a wide regime emerging in $Ru_{1-x}Cr_xO_2$ (x ≤ 0.2) should be ascribed to the unique orbital reconstruction between Cr and Ru. We envision that this design strategy can be extended to more systems to produce further exotic phenomena.

## Methods

**DFT calculations and Wannierization.** We computed the Bloch wavefunctions for $RuO_2$ on the basis of density functional theory (DFT) using the Quantum ESPRESSO package [41,42]. We first assumed a nonmagnetic structure without spin-orbit coupling and used the projector augmented wave pseudopotential [43] and the generalized gradient approximation of the Perdew-Burke-Ernzerhof exchange correlation functional [44]. We used lattice constants of a = 4.492 Å and c = 3.107 Å. The energy cutoff for the wave function and the charge density, $e_{wfc}$ and $e_{rho}$, respectively, were set to $e_{wfc}$ = 60 Ry and $e_{rho}$ = 400 Ry. We used **k**-point meshes of 12×12×16 and 16×16×16 in the self-consistent field (scf) and non-scf calculations, respectively. After the DFT calculations, Wannierization was performed by using the wannier90 package [45,46], in which the Bloch orbitals were projected onto the $t_{2g}$ orbitals of Ru ions with 16×16×16 **k**-point grids.

To calculate the electronic states of $Ru_{1-x}Cr_xO_2$, with x = 0, 0.25, and 0.5, we replaced the Ru-sites denoted as Ru-1 or Ru-2 in Supplementary Fig. 3a with Cr. In this calculation, we set $e_{rho}$ = 500 Ry, and the spin-orbit coupling was not included. For the x = 0 and 0.5 systems, we took 24×24×32 **k**-mesh for the scf calculation. When we calculated the ground states of $Ru_{0.75}Cr_{0.25}O_2$, we used the supercell with the b- or c-axis doubled. We took the **k**-mesh of 24×12×32 (24×24×16) when the b- (c-)axis was doubled for the scf calculation. We found that the supercell with the b-axis doubled was more energetically stable, which we have used for discussion. To obtain the projected density of states (PDOS) of the x = 0 and 0.5 systems, we performed the non-scf calculations with 24×24×32 $k$-mesh after the scf calculation and then calculated the PDOS. We also calculated the PDOS of $RuO_2$ with the DFT+$U$ method with $U = 3$ eV and nonmagnetic $Ru_{1-x}Cr_xO_2$ with x = 0 and 0.5, where we set $e_{rho}$ = 500 Ry and took 24×24×32 $k$-points for the scf and non-scf calculations.

For examining the orientation of the Néel vector, we performed the DFT+$U$ calculation for $RuO_2$ with the spin-orbit coupling for the three cases where the Néel vector was initially along [001], [100], and [110]. We took $U = 3$ eV. We used 24×24×32 $k$-points and set $e_{rho}$ = 500 Ry. The convergence threshold for the calculation of the Néel vector orientation was set as $10^{-6}$ Ry.

**DMFT calculations.** The Wannier functions obtained above define a tight-binding model for the three Ru $t_{2g}$ orbitals of $RuO_2$. Using this as the one-body part of the Hamiltonian, we constructed a multiorbital Hubbard model with intra(inter)orbital Coulomb interaction $U(U')$ and Hund's coupling and pair hopping $J$. We solved the model within the dynamical mean field theory (DMFT) [47] at zero temperature. As a solver for the DMFT impurity problem, we used the exact diagonalization method [48], where the dynamical mean field was represented by nine bath sites. To obtain the antiferromagnetic solution, we assumed opposite spin polarizations at neighboring Ru sites in the unit cell. For the interaction parameters, we assumed $U = U' + 2J$ and $J = U/5$ for the sake of simplicity.

**Thin-film growth, X-ray diffraction, and XAS.** The $Ru_{1-x}Cr_xO_2$ films were grown on the rutile $TiO_2$ substrate by the PLD method with stoichiometric targets. During sample growth, the substrate temperature was kept at 290 °C to suppress interfacial diffusion, and the oxygen partial pressure was kept at 20 mTorr. The laser fluence was 1.2 J/cm² (KrF, λ= 248 nm), and the deposition frequency was 3 Hz. After deposition, the samples were cooled to room temperature at a rate of 10 °C/min under an oxygen pressure of 10 Torr. The film thickness was determined directly with an X-ray reflectivity measurement. X-ray diffraction measurements were performed using a high-resolution diffractometer (Rigaku) with monochromatic Cu K$_{α1}$ (λ = 1.5406 Å) X-rays. The stoichiometry in the thin film was checked by energy dispersive X-ray (EDX), and the ratio of Ru/Cr was confirmed to be very close to the target. The XAS curves of Cr L-edge were measured with a total electron mode, at 20 K, in beamline BL07U of Shanghai Synchrotron Radiation Facility.

**Transport and magnetization measurements.** All of the electrical transport was carried out on Hall bar devices with a size of 300 μm × 60 μm, which were fabricated by photolithography. The milling process was carried out with $Ar/O_2$ (10:1) mixed ions and at a low speed to avoid oxygen vacancy formation on the $TiO_2$ surface. The transport measurements were carried out with a PPMS system (Quantum Design) with an in-plane DC current. The magnetoresistivity (MR) and its anisotropy were very small, as shown in Supplementary Fig. 10. The Hall conductivity $\sigma_{xy}$ was calculated as $\sigma_{xy} = -\rho_{yx}/(\rho_{xx}^2)$. The magnetization was measured using an MPMS system (Quantum Design) and obtained by subtracting the contribution from the $TiO_2$ substrate. The Hall vector ($\sigma_{Hall}$) is defined as that in reference 14.

**Acknowledgments:** This research was supported by JSPS KAKENHI (Grant Nos. 21H04437, 19H05825, 19H02594, 21H04442, 21K14398) and JST CREST (Grant No. JPMJCR1874). P. Y. was financially supported by the National Key R&D Program of China (grant No. 2021YFE0107900) and the National Natural Science Foundation of China (grant No. 52025024).

**Author Contributions:** M. W. and F. K. conceived the project. M. W. grew the thin films and performed the transport measurements with help from S. Shen. K. T. and S. Sakai performed the calculations with the supervision of R. A. M.W., Y. L., and D. T. conducted the XRD and magnetization measurements with support from P. Y. K. D. and C. L. conducted the XAS measurements. Z. W. and N. O. performed the Kerr mapping. M. W. and F. K. wrote the manuscript. All of the authors discussed the results and provided feedback.

**Competing Interests:** The authors declare no competing interests.

**Data availability:** All data used to generate the figures in the manuscript and supplementary information is available on Zenodo at: https://zenodo.org/record/8128770

# References

1. Miyasato, T. et al. Crossover behavior of the anomalous Hall effect and anomalous Nernst effect in itinerant ferromagnets. *Phys. Rev. Lett*. **99**, 086602 (2007).

2. Nagaosa, N., Sinova, J., Onoda, S., MacDonald, A. H. & Ong, N. P. Anomalous Hall effect. *Rev. Mod. Phys*. **82**, 1539-1592 (2010).

3. M.-T. Suzuki, T. Koretsune, M. Ochi, R. Arita, Cluster multipole theory for anomalous Hall effect in antiferromagnets. *Phys. Rev. B* **95**, 094406 (2017).

4. Chen, H., Niu, Q. & MacDonald, A. H. Anomalous Hall effect arising from noncollinear antiferromagnetism. *Phys. Rev. Lett*. **112**, 017205 (2014).

5. M. Tinkham, Group theory and quantum mechanics. (McGraw-Hill, Inc., 1964), pp.229-309.

6. Nakatsui, S., Kiyohara, N. & Higo, T. Large anomalous Hall effect in a non-collinear antiferromagnet at room temperature. *Nature* **527**, 212-215 (2015).

7. Nayak, A. K. et al. Large anomalous Hall effect driven by a nonvanishing Berry curvature in the noncolinear antiferromagnet $Mn_3Ge$. *Sci. Adv*. **2**, e1501870 (2016).

8. Kim, W. J. et al. Strain engineering of the magnetic multipole moments and anomalous Hall effect in pyrochlore iridate thin films. *Sci. Adv*. **6**, eabb1539 (2020).

9. Ghimire, N. J. et al. Large anomalous Hall effect in the chiral-lattice antiferromagnet $CoNb_3S_6$. *Nat. Commun*. **9**, 3280 (2018).

10. Liu, X. et al. Magnetic Weyl semimetallic phase in thin films of $Eu_2Ir_2O_7$. *Phys. Rev. Lett*. **127**, 277204 (2021).

11. Suzuki, T. et al. Large anomalous Hall effect in a half-Heusler antiferromagnet. *Nat. Phys*. **12**, 1119-1123 (2016).

12. Suzuki, M.-T. et al., Multipole expansion for magnetic structures: A generation scheme for a symmetry-adapted orthonormal basis set in the crystallographic point group. *Phys. Rev. B* **99**, 174407 (2019).

13. Seemann, M., Ködderitzsch, D., Wimmer, S. & Ebert, H. Symmetry-imposed shape of linear response tensors. *Phys. Rev. B* **92**, 155138 (2015).

14. Šmejkal, L., González-Hernández, R., Jungwirth, T. & Sinova, J. Crystal time-reversal symmetry breaking and spontaneous Hall effect in collinear antiferromagnets. *Sci. Adv*. **6**, eaaz8809 (2020).

15. Berlijn, T. et al. Itinerant antiferromagnetism in $RuO_2$. *Phys. Rev. Lett*. **118**, 077201 (2017).

16. Šmejkal, L., MacDonald, A.H., Sinova, J. et al. Anomalous Hall antiferromagnets. *Nat. Rev. Mater.* **7**, 482-496 (2022).

17. Feng, Z. et al. An anomalous Hall effect in altermagnetic ruthenium dioxide. *Nat. Electron*. **5**, 735-743 (2022).
18. Feng, Z. et al. Observation of the anomalous Hall effect in a collinear antiferromagnet. https://arxiv.org/abs/2002.08712.
19. Zhong, D. et al. Layer-resolved magnetic proximity effect in van der Waals heterostructures. *Nat. Nano*. **15**, 187-191 (2020).
20. Williams, A. J. et al. Charge transfer and antiferromagnetic insulator phase in $SrRu_{1-x}Cr_xO_3$ perovskites: Solid solutions between two itinerant electron oxides. *Phys. Rev. B* **73**, 104409 (2006).
21. Korotin, M. A., Anisimov, V. I., Khomskii, D. I. & Sawatzky, G. A. $CrO_2$: A self-doped double exchange ferromagnet. *Phys. Rev. Lett*. **80**, 4305 (1998).
22. Karube, S. et al. Observation of spin-splitter torque in collinear antiferromagnetic $RuO_2$. *Phys. Rev. Lett*. **129**, 137201 (2022).
23. Zhang, K. H. L. et al. Hole-induced insulator-to-metal transition in $La_{1-x}Sr_xCrO_3$ epitaxial films. *Phys. Rev. B* **91**, 155129 (2015).
24. Mugiraneza, S. & Hallas, A. M. Tutorial: A beginner's guide to interpreting magnetic susceptibility data with the Curie-Weiss law. *Commun. Phys*. **5**, 1-12 (2022).
25. Blundell, S. Magnetism in Condensed Matter (Oxford University Press New York, 2001).
26. Zhou, J.-S., Jin, C.-Q., Long, Y.-W., Yang, L.-X. & Goodenough, J. B. Anomalous electronic state in $CaCrO_3$ and $SrCrO_3$. *Phys. Rev. Lett*. **96**, 046408 (2006).
27. Schmehr, J. L. et al. Overdamped antiferromagnetic strange metal state in $Sr_3IrRuO_7$. *Phys. Rev. Lett*. **96**, 046408 (2019).
28. Peng, H. et al. A generic sacrificial layer for wide-range freestanding oxides with modulated magnetic anisotropy. *Adv. Funct. Mater*. **32**, 2111907 (2022).
29. Koster, G. et al. Structure, physical properties, and applications of $SrRuO_3$ thin films. *Rev. Mod. Phys*. **84**, 253-298 (2012).
30. Shim, J. H., Lee, S., Dho, J. & Kim, D.-H. Coexistence of two different Cr ions by self-doping in half-metallic $CrO_2$ Nanorods. *Phys. Rev. Lett*. **99**, 057209 (2007).
31. Bartram, F. M. et al. Anomalous Kerr effect in $SrRuO_3$ thin films. *Phy. Rev. B* 102, 140408(R) (2020).
32. Greedan, J. E. Geometrically frustrated magnetic materials. *J. Mater. Chem*. 11, 37 (2001).
33. Miyasato, T. et al. Crossover behavior of the anomalous Hall effect and anomalous Nernst effect in itinerant ferromagnets. *Phys. Rev. Lett*. **99**, 086602 (2007).
34. Fujishiro, Y. et al. Giant anomalous Hall effect from spin-chirality scattering in a chiral magnet. *Nat. Commun*. **12**, 317 (2021).
35. Tian, Y., Ye, L., & Jin, X. Proper scaling of the anomalous Hall effect. *Phys. Rev. Lett*.

**103**, 87206 (2009).

36. Li, Y. F. et al. Robust formation of skyrmions and topological Hall effect anomaly in epitaxial thin films of MnSi. *Phys. Rev. Lett*. **110**, 117202 (2013).

37. Wang, L. et al. Controllable thickness inhomogeneity and Berry curvature engineering of anomalous Hall effect in SrRuO$_3$ ultrathin films. *Nano. Lett*. **20**, 2468−2477 (2020).

38. Tian, D. et al. Manipulating Berry curvature of SrRuO$_3$ thin films via epitaxial strain. *PNAS*, **118**, (18) e2101946118 (2021).

39. Gonzalez Betancourt, R. D. et al. Spontaneous anomalous Hall effect arising from an unconventional compensated magnetic phase in a semiconductor. *Phys. Rev. Lett* **130**, 036702 (2023).

40. Song, Q. et al. Antiferromagnetic metal phase in an electron-doped rare-earth nickelate. *Nat. Phys*. **19**, 522-528 (2023).

41. Giannozzi, P. et al. QUANTUM ESPRESSO: a modular and open-source software project for quantum simulations of materials. *J. Phys. Condens. Matter* **21**, 395502 (2009).

42. Giannozzi, P. et al. Advanced capabilities for materials modelling with QUANTUM ESPRESSO. *J. Phys. Condens. Matter* **29**, 465901 (2017).

43. Blöchl P. E. Projector augmented-wave method. *Phys. Rev. B* **50**, 17953-17979 (1994).

44. Perdew, J. P., Burke, K. & Ernzerhof, M. Generalized gradient approximation made simple. *Phys. Rev. Lett.* **77**, 3865-3868 (1996).

45. Mostofi, A. A. et al. Wannier90: A tool for obtaining maximally-localised Wannier functions. *Comput. Phys. Commun*. **178**, 685-699 (2008).

46. Pizzi, G. et al. Wannier90 as a community code: New features and applications. *J. Phys.Condens. Matter* **32**, 165902 (2020).

47. Georges, A., Kotliar, G., Krauth, W. & Rozenberg, M. J. Dynamical mean-field theory of strongly correlated fermion systems and the limit of infinite dimensions. *Rev. Mod. Phys*. **68**, 13 (1996).

48. Caffarel, M. & Krauth, W. Exact diagonalization approach to correlated fermions in infinite dimensions: Mott transition and superconductivity. *Phys. Rev. Lett*. **72**, 1545 (1994).

# Figures

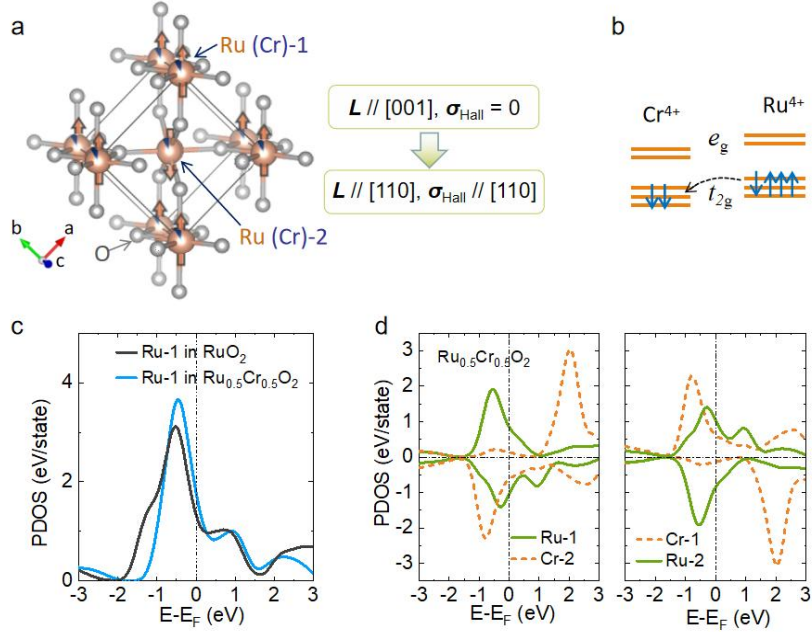

**Fig. 1 Antiferromagnetic symmetry controlled anomalous Hall effect (AHE) and DFT+DMFT calculations for Cr-doped $RuO_2$. a**, Crystal structure of the Cr-doped rutile phase $RuO_2$. O-ions are located between two Ru (Cr) sites asymmetrically. The Ru-1 (Cr-1) and Ru-2 (Cr-2) denote the Ru (Cr) ions at the center and the corner sites of the unit cell, respectively. The orange arrows denote the local magnetic moment with antiferromagnetic coupling along [110]. Hall vector ($\sigma_{Hall}$) is allowed and parallel to the Néel vector (**L**) along [110] in such a configuration, which vanishes as the Néel vector is along [001], indicating a manipulating of **L** is necessary to generate AHE. **b**, Schematic illustration of charge transfer in Cr-doped $RuO_2$. The orbital level difference between the nearest neighbor sites can lead to partial charge transfer from $Ru^{4+}$ to $Cr^{4+}$ to form a reconstructed Fermi level and maintain an antiparallel spin coupling. **c**, Calculated projected density of states (PDOS) of the $RuO_2$ and $Ru_{0.5}Cr_{0.5}O_2$ in the paramagnetic phase. The Ru-2 sites for both components possess identical PDOS with Ru-1. **d**, Calculated PDOS of the $Ru_{0.5}Cr_{0.5}O_2$ in the magnetic ground state. The doped Cr ions have two selective sites as labeled by Cr-1 and Cr-2 in (a). Ru and Cr both show an asymmetric PDOS (a spontaneous

polarization), while exhibiting an antiparallel coupling.

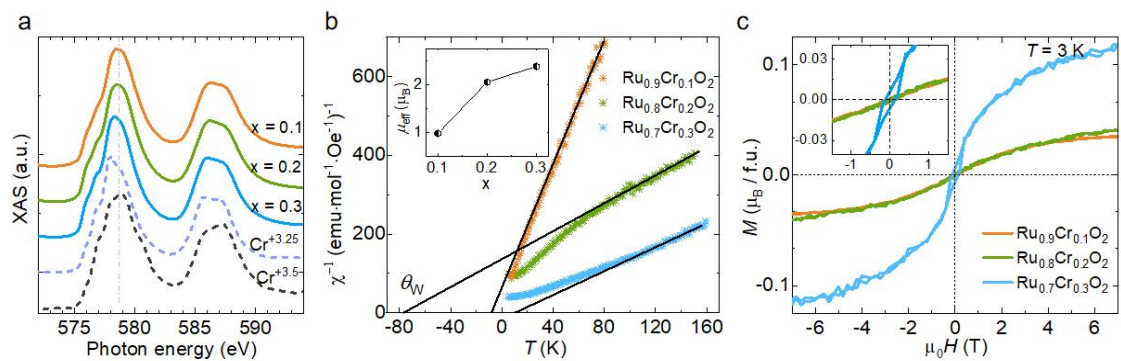

**Fig. 2 XAS and magnetic states evolution in $Ru_{1-x}Cr_xO_2$ films grown on $TiO_2$ (110). a**, XAS around the *L*-edge of Cr measured in the $Ru_{1-x}Cr_xO_2$ films compared to that in $La_{0.75}Sr_{0.25}CrO_3$.[23] **b,c**, Temperature-dependent magnetic susceptibility (b) and magnetic field dependent magnetization (c) curves measured with a magnetic field along the out-of-plane (OOP) axis. All films are grown on $TiO_2$ (110). Inset of (b), the effective on-site moments ($\mu_{eff}$) depending on the doping level x. Inset of (c), an expanded view of the low-field region. Linear fittings of the $\chi^{-1}$-*T* curves at high temperatures indicate an antiferromagnetic behavior with negative Weiss temperatures $\theta_W$ = -10 K and -75 K in x = 0.1 and 0.2, respectively. The x = 0.3 film shows a small positive $\theta_W$, with a finite remanent magnetization at zero field (0.008 $\mu_B$/f.u.), implying a ferrimagnetic ground state. The magnetic field is 1 T for the $\chi^{-1}$-*T* measurement.

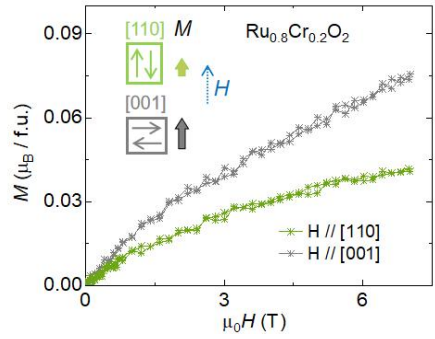

**Fig. 3 Magnetic anisotropy and Néel vector orientation in $Ru_{0.8}Cr_{0.2}O_2$ film grown on $TiO_2$ (110).** *M-H* curves were measured at 3 K with magnetic field along out-of-plane (*H* ∥ [110]) and in-plane (*H* ∥ [001]), respectively. Inset, the illustration of spin orientation and corresponding canting moments (wide arrows) driven by the external magnetic field (*H*) applied to [110] and [001].

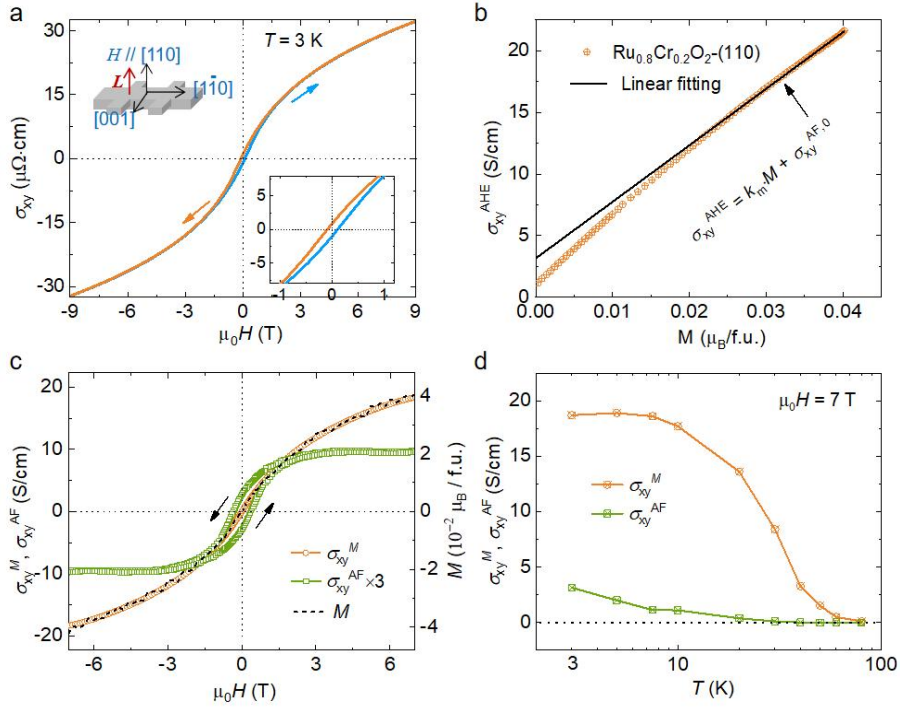

**Fig. 4 Transport properties of the Ru$_{0.8}$Cr$_{0.2}$O$_2$ film grown on TiO$_2$ (110). a**, Hall conductivity with magnetic field dependence at 3 K. Insets show the Hall configuration (left) and an expanded view of the low-field region (right). **b**, Anomalous Hall conductivity at 3 K with a dependence on the magnetic moment (*M*). The σ$_{xy}$$^{AHE}$ was obtained by subtracting a field-linear-dependent ordinary Hall contribution from σ$_{xy}$. *M* was measured by an MPMS at 3 K. **c**, Anomalous Hall conductivity derived from the canting moment (i.e., σ$_{xy}$$^M$) and the antiferromagnetic domain (i.e., σ$_{xy}$$^{AF}$) in Ru$_{0.8}$Cr$_{0.2}$O$_2$ (110) with a dependence on magnetic field sweeping at 3 K. The magnetic moment is shown by a dashed line. **d**, Temperature-dependent σ$_{xy}$$^M$ and σ$_{xy}$$^{AF}$. The data at 7 T are used.

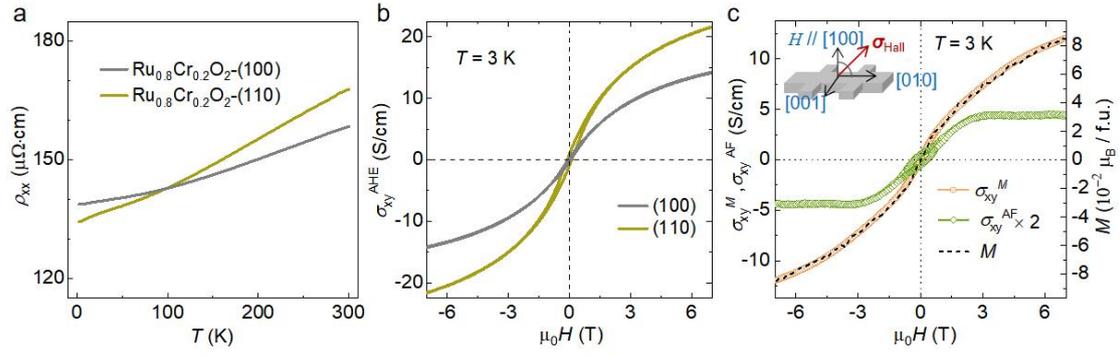

**Fig. 5 Comparison of the transport behaviors for films grown along (100) and (110). a,** Temperature-dependent longitudinal resistivities of the two films. **b,** Magnetic field-dependent $\sigma_{xy}^{AHE}$ at 3 K for the two films. **c,** $\sigma_{xy}^{M}$ and $\sigma_{xy}^{AF}$ with magnetic-field dependence at 3 K for the film grown along the (100) orientation. The magnetic moment is shown by a dashed line. During the transport measurement on the $Ru_{0.8}Cr_{0.2}O_2$-(100) film, the current was applied along the [010] direction with a Hall voltage along the [001] direction for comparison. Inset, the illustration of the Hall bar and the Hall vector ($\boldsymbol{\sigma}_{Hall}$).